\documentclass[3p,times,procedia]{elsarticle}
\usepackage{nupha_ecrc}
\usepackage{hyperref}

\volume{00}

\firstpage{1}

\journalname{Nuclear Physics A}

\runauth{C.Loizides}

\jid{nupha}

\jnltitlelogo{Nuclear Physics A}

\usepackage{amssymb}
\usepackage{amsthm}

\biboptions{sort&compress}

\usepackage[figuresright]{rotating}

\newif\ifcomment
\newif\ifarxiv
\arxivtrue
\newcommand{\pp}          {pp}
\newcommand{\ep}          {ep}
\newcommand{\eA}          {eA}

\renewcommand{\AA}       {AA}

\newcommand{\pA}          {pA}

\newcommand{\pPb}         {pPb}

\newcommand{\AuAu}      {AuAu}
\newcommand{\ArAr}       {ArAr}
\newcommand{\OO}         {OO}
\newcommand{\PbPb}       {PbPb}

\newcommand{\mub}       {\ensuremath{\mu_{\rm B}}}
\newcommand{\pT}           {\ensuremath{p_{\rm T}}}
\newcommand{\pt}            {\pT}
\newcommand{\RAA}         {\ensuremath{R_{\rm AA}}}

\newcommand{\RpPb}        {\ensuremath{R_{\rm pPb}}}

\newcommand{\Npart}       {\ensuremath{N_{\rm part}}}

\newcommand{\snn}          {\ensuremath{\sqrt{s_{\rm NN}}}}
\newcommand{\s}              {\ensuremath{\sqrt{s}}}

\newcommand{\hrefurl}[1]  {\href{#1}{\url{#1}}}

\newcommand{\Fig}[1]       {Fig.~\ref{#1}}
\newcommand{\Figure}[1]  {Figure~\ref{#1}}

\newcommand{\Sec}[1]       {Sec.~\ref{#1}}

\newcommand{\com}[1]      {}

\begin{document}

\begin{frontmatter}



\dochead{XXVIIIth International Conference on Ultrarelativistic Nucleus-Nucleus Collisions\\ (Quark Matter 2019)}

\title{``QM19 summary talk'':\\Outlook and future of heavy-ion collisions}


\author{Constantin Loizides}

\address{ORNL, Oak Ridge, USA}

\begin{abstract}
A summary of the QM19 conference is given by highlighting a few selected results.
These are discussed as examples to illustrate the exciting future of heavy-ion collisions and the need for further instrumentation.
\ifarxiv
(The arXiv version is significantly longer than the printed proceedings, with more figures.)
\fi
\end{abstract}




\end{frontmatter}


\section{Introduction}
\label{sec:intro}
The goal of heavy-ion physics~\cite{Busza:2018rrf} is to understand the phase diagram of Quantum Chromo Dynamics~(QCD) as a function of the temperature~($T$) and baryon chemical potential~($\mub$)\ifarxiv, as shown in \Fig{fig:phasediagram}\fi.
At high temperature and/or high baryon chemical a transition occurs, from ordinary matter, where the hadronic degrees of freedom are dominant, to a Quark Gluon Plasma~(QGP), where the dominant degrees of freedom are quark and gluons, which in ordinary matter are confined into hadrons.
Lattice QCD calculations~\cite{Ding:2020rtq} predict the presence of a Critical Endpoint~(CE) somewhere in the region $T<140, \mub>300$~MeV with a first-order phase transition at higher $\mub$ and a cross-over transition at $T\approx 155$~MeV and lower $\mub$. 
Immense experimental and theoretical effort is underway to characterize and understand the phase structure of QCD, and the emergence of collectivity and matter properties, and the underlying equation of state~(EoS) from first principles.
Following the presentation at the conference~\cite{Loizides:qm19}, these proceedings are structured into the following topics: 
\ifarxiv
\begin{itemize}
\item the high-density frontier~(\Sec{sec:highdensity}) to study the onset of deconfinement, to search for the CE and the first-order phase transition, and at very high $\mub$ to provide constraints for the  neutron star structure and its EoS;
\item the high-energy frontier~(\Sec{sec:highenergy}) to quantify the fluid properties of QGP and to relate them to its microscopic structure;
\item the cold nuclear matter or small-$x$ frontier~(\Sec{sec:coldmatter}) to characterize properties of cold nuclear matter, and understand the structure of protons and nuclei at small Bjorken-$x$;
\item the ultra-precision near and far future~(\Sec{sec:future}) with the planned Electron Ion Collider~(EIC) and the proposed new future experimental equipment at the Large Hadron Collider~(LHC) and beyond.
\end{itemize}
\fi

\ifarxiv
\begin{figure}[t!]
    \centering
    \includegraphics[width=0.6\textwidth]{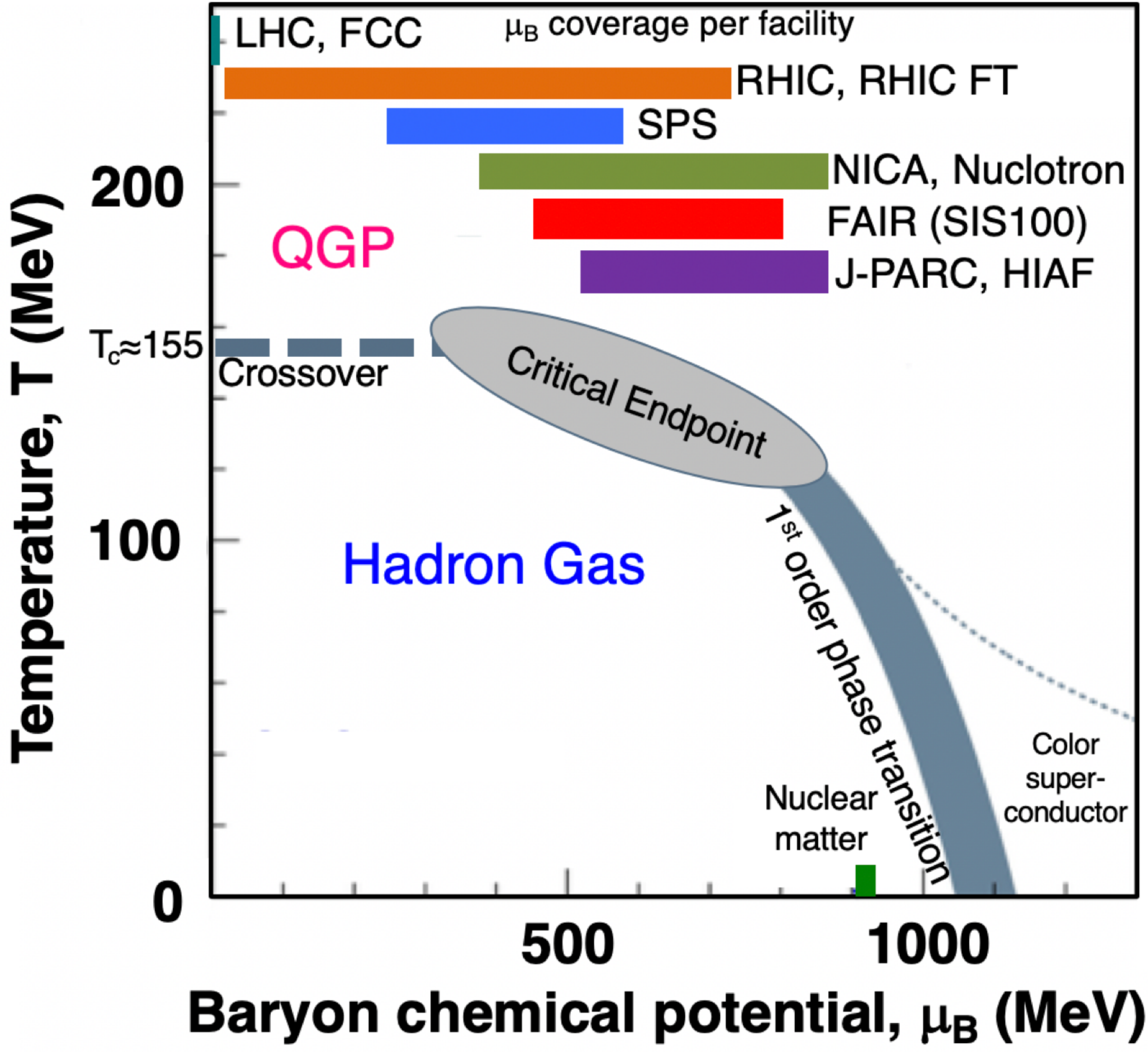}
    \caption{
      The QCD phase diagram versus temperature and bayon chemical potential.
      Indicated are the nuclear matter, hadron gas and QGP phases, as well as the approximate region for the Critical Endpoint, and the first-order and cross-over phase transitions. 
      Current and future facilities probing different region in $\mub$ are: LHC~\cite{Evans:2008zzb}, FCC~\cite{Benedikt:2020ejr}, RHIC~\cite{Harrison:2003sb}, and the RHIC fixed-target program~\cite{Aggarwal:2010cw}, SPS~\cite{Aduszkiewicz:2309890,Dahms:2673280}, NICA~\cite{Kekelidze:2017ghu}, FAIR~\cite{Senger:2020pzs} and J-PARC~\cite{Sako:2015cqa}.
      Figure from \cite{Dainese:qm19}.}
    \label{fig:phasediagram}
\end{figure}
\fi
  
\section{The high-density frontier}
\label{sec:highdensity}
Predictions for the existence and exact location of the CE at finite $\mub$ have been continuously improved over the past years.
New lattice QCD results from the WB collaboration~\cite{Borsanyi:2020fev} using the imaginary $\mub$ method are consistent with previous calculations by the HotQCD collaboration, but have much smaller uncertainty, and hence add further evidence that the location of CP is disfavoured for $\mub<300$~MeV.
Where possible, lattice QCD predictions were already confronted with experimental data~\cite{Bzdak:2019pkr}.
The experimental approach~\cite{Andronic:2017pug} is to probe different regions of the QCD matter phase diagram by extracting the freeze-out parameters~($T$, $\mub$) from statistical model fits to hadron yields measured at LHC~(few TeV) down to SIS18~(few GeV) collision energies.
In this way, the vicinity of the phase boundary will be explored from $T\approx155$ and $\mub\approx0$~MeV down to about $T=80$ and $\mub=900$~MeV\ifarxiv as shown in \Fig{fig:phasediagram}\fi.

The CE is characterized as a divergence in the correlation length of the underlying system, and hence manifests as a divergence of the associated susceptibilities.
Experimentally, ratios of susceptibilities are accessible through event-by-event fluctuations in conserved quantities, such as electric charge or baryon number.
In particular, higher-order moments of net-proton distributions, which are a proxy for baryon number, are expected to be sensitive to the CE.
First results of $C_6/C_2$ of net-proton distributions were reported~\cite{Nonaka:2020crv}, which for central \AuAu\ collisions at $\snn=200$~GeV are negative, as expected from lattice QCD, while they were found to be positive for central \AuAu\ collisions at $\snn=54.4$~GeV.
Albeit being-only a $2\sigma$-effect at present, the result may indicate the expected $O(4)$ criticality in the cross-over region~\cite{Friman:2011pf}.
Furthermore, a crucial study was performed~\cite{Gupta:2020pjd} by checking that the freeze-out parameters deduced from mean hadron yields agree with those from grand-canonical fluctuations of conserved quantities for central collisions.
This demonstrates for the first time, using fluctuation observables, that a femto-scale system attains thermalization.
A low $\snn$~($<19.6$~GeV) the presented study however indicates that the created system may be too dilute to reach equilibrium or exhibits different relaxation times for different moments.
In view of the search for the CE this should be followed-up on. 

A first-order phase transition is characterized by an unstable co-existence~(spinodal) region corresponding to a softest point in the EoS.
Direct flow~($v_1$) which is sensitive to the compressibility of the created matter, is hence a key observable in the beam energy scan~(BES) program.
The slope of the rapidity-odd component at mid-rapidity was reported as a function of beam energy for various identified mesons and baryons~\cite{Nayak:2020djj}.
Mesons and anti-baryons exhibit negative slope over the whole range of beam energies, while baryons~(p, $\Lambda)$ exhibit a change of slope around $14.5$~GeV.
In hydrodynamical models a change of slope implies a minimum in $v_1$, and hence was proposed as a signature of a first-order phase transition~\cite{Stoecker:2004qu}.
However, to consistently treat effects from spectator matter and baryon stopping, calculations in the frame of the BEST collaboration~\cite{best} would be needed using more realistic 3D hydrodynamic calculations at finite $\mub$~\cite{Shen:2020gef},
with an EoS with and without CE~\cite{Parotto:2018pwx}.
First data from the STAR fixed-target~(FT) program, the $v_1$  of the $\phi$-meson at $\snn=4.5$~GeV were also shown, but the statistical uncertainties are too large to allow for additional conclusions.
More results, from BES phase 2, and the fixed-target program, are eagerly awaited.

\section{The high-energy frontier}
\label{sec:highenergy}
The high-energy experiments at RHIC and LHC continue to produce data with ever increasing precision in measuring bulk properties.
At the conference, for example, new results of longitudinal flow decorrelations at LHC~\cite{Aad:2020gfz} and their energy dependence at RHIC~\cite{Nie:2020trj}, linear and non-linear flow modes~\cite{Acharya:2019uia} and higher order cumulant elliptic flow and fluctuations~\cite{YaZhu} of identified particles in \PbPb\ collisions at $\snn=5.02$ TeV were presented.
These and many more data are now rigorously used in Bayesian extraction of bulk properties, for example using the JETSCAPE framework~\cite{Putschke:2019yrg}.
It was argued~\cite{Paquet:2020rxl} that RHIC and LHC data add complementary constraints, and that there are non-negligible uncertainties in the transition from fluid cells to particles (“viscous corrections”).
In developing community frameworks like JETSCAPE, it is important to highlight and maintain  the``plug-and-play'' approach of the framework, which allows one to interchange and study different descriptions of the various stages of the collision.
Furthermore, we need to insist that contributing authors release their pieces of code as open source promptly.

The precision reached in the J$/\psi$ nuclear modification factor at mid-rapidity from ALICE~\cite{Bai} is now good enough to clearly see predicted behavior of the $\RAA$ from statistical hadronization or regeneration~\cite{Gazdzicki:1999rk,Thews:2000rj,Andronic:2003zv} counteracting Debye screening:
A clear minimum is observed in peripheral collisions~($\Npart\sim100$), beyond which the $\RAA$ rises for more central collisions almost reaching unity in central collisions.
Consequently, one also observes large flow as well as triangular flow due to ``approximately thermalized'' charm at low $\pT$~\cite{Acharya:2020jil}.
The total charm cross section needs to be measured, since it is the natural normalization for J/$\psi$ measurements.
Knowing it precisely allows one to reduce model uncertainties by constraining shadowing effects, and set precise limits on the expected $\RAA$.
Unity is not apriori a limiting value; the J/$\psi$ $\RAA$ can be larger than one~\cite{Andronic:2017pug}
With a ``brute-force'' combinatorial extraction of D meson production down to 0.5~GeV/$c$, first constraints of the total charm cross section in \PbPb\ were placed~\cite{Innocenti}.
Better performance and more statistics will be expected in Run-3 after the LS2 ALICE upgrades. 

The situation for the $Y$ is different as due to its much heavier mass one does not expect a significant regeneration component.
Broadening of its spectral widths on the lattice for finite temperature, presented at the conference~\cite{Larsen:2019zqv}, is compatible with the sequential dissociation picture.
Sequential dissociation is clearly established for the $Y$ family, and observed to be significantly stronger in \PbPb\ than \pPb\ collisions~\cite{1801111}.
In \pPb\ collisions, the suppression usually is attributed to comoving nuclear matter, but could in principle also be partially from the onset of screening.
Elliptic flow was found to be consistent with zero~\cite{CMS-PAS-HIN-19-002,Acharya:2019hlv}, however from extrapolating Blast Wave fits of existing data, one expects significant $v_2$ only for $\pT>10$ GeV~\cite{Reygers:2019aul}.

Understanding and characterizing parton energy loss continues to evolve from rather qualitative to more-and-more sophisticated and precise level.
A milestone was reached with the first measurement of the jet fragmentation of jets recoiling of an isolated photon~\cite{Sirunyan:2018qec}.
The data clearly exhibit the expected ``text book'' result that soft (low-$z$) particles are enhanced, while hard (high-$z$) are suppressed, as predicted since more than 20 years~(e.g.see \cite{Borghini:2005em} and references therein).
Detailed measurements using isolated photon--jet correlations will play a key role to precisely quantify jet modification in Run-3/4 at the LHC~\footnote{At LHC also Z--jet correlations are becoming available~\cite{CMS-PAS-HIN-19-006}} and in the sPHENIX era at RHIC.
Measurements of inclusive jet $\RAA$ were performed from about 50 to 500 GeV/$c$ for $R=0.4$ in \PbPb\ collisions at $\snn=5.02$ TeV, and rather good agreement was found between data and models~\cite{Aaboud:2018twu,Acharya:2019jyg}.
It was hence surprising that the predictions~\cite{CMS:2019btm} for larger radii and to higher $\pt$ vary considerably indicating that the limited available measurements in the past year attracted the calculations.
In order to avoid tuning of models to specific results, we could define a set of reference measurements~(accord) that every jet model has to be able to describe before one trusts its prediction for a different observable.
The CMS collaboration released new results~\cite{CMS:2019btm} of $\RAA$ for high energy jets up to $R=1.0$, which yield $\RAA\approx0.8$ above 300~GeV/$c$ for central \PbPb\ collisions at $\snn=5.02$~TeV.
Peripheral collisions are consistent with unity, in particular when taking into account also the presence of the centrality bias~\cite{Morsch:2017brb}.
The jet suppression for large radii in central collisions is similar to that of charged particles at $\sim200$~GeV/$c$~\cite{Khachatryan:2016odn}, indicating the sought-after consistency between inclusive charged particle and jet $\RAA$ at high $\pt$ and large enough radius.
Still, with new data from Run-3/4, it is important to measure both to even higher $\pT$, to reduce uncertainties, and to investigate if $\RAA\sim1$ will be reached.
\ifarxiv
Further progress with jet measurements to lower $\pt$ and large radii~($\ge0.4$) can be made by using Machine Learning methods trained on properties of the jet constituents in \pp\ collisions~\cite{Haake:2018hqn}.
First results have already been presented\cite{Haake:2019pqd}, and confirm the previously established picture of strong quenching at low jet $\pT$.
However, in particular for large jet radii of $R=0.6$ a potential bias induced by changed jet fragmentation needs to be investigated. 
\fi

Substructure and reclustering techniques are extremely promising window to access the partonic structure of the shower~\cite{Marzani:2019hun}.
New evidence for the coherent/wide angle energy loss picture have been presented at the conference.
Symmetric splittings for wide-angles were reported to be suppressed relative to vacuum~\cite{Havener:2020ett} and single sub-jets less exhibit less suppression than jets with multiple subjets~\cite{ATLAS-CONF-2019-056}.

One of the frontiers of the high-energy program is to study the evolution and onset of QGP phenomena in smaller collision systems like \pA\ and \pp\ collisions~\cite{Nagle:2018nvi}. 
The yields of strange and multi-strange particles~(relative to pions) increase smoothly with multiplicity, rather independent of collision species and energy~\cite{pisano} reaching a value predicted by statistical models in central \PbPb\ collisions.
The increase with multiplicity in smaller systems can be regarded as lifting of canonical suppression~\cite{Vislavicius:2016rwi}.
The rapid rise at low multiplicity plus the presence of many QGP signatures in \pp\ and \pPb\ collisions~\cite{Loizides:2016tew} lead to the development of ``core/corona'' models.
In these models, the core ``hydrodynamizes'', while the corona does not and instead is treated as a superposition of independent nucleon--nucleon collision.
In this way, good description of data across all systems is achieved with a rather universal approach~(similar to EPOS)~\cite{Kanakubo:2019ogh}.

The observed enhancement in the strange particle yields versus multiplicity result from the fact that there is a pedestal effect:
The strange particle yields depend approximately linearly on multiplicity but only above some minimum multiplicity~(threshold), while the pion yields do not exhibit any apparent threshold.
Hence, normalizing the strange particle yields with the pion yields creates an enhancement $\propto (1-M_0/M)$, which leads to a rise with $M$ that is most apparent at low multiplicities.
In microscopic models one can interpret the observed threshold in several ways:
i)~The yield increases with the number of overlapping strings, i.e.\ as a multi-string effect related to the string density driving the increase of strangeness production that however quickly saturates at higher multiplicity.
ii)~There is minimum associated multiplicity necessary for multi-strange particle production and/or different scaling of soft and (semi-)hard processes, i.e.\ a single string effect leading to suppression at low multiplicity that is reduced at higher multiplicity.
To further probe production mechanism of strangeness can be achieved by measuring two-particle angular correlations~(associated production) between same and opposite sign strange and non-strange particles~\cite{Adolfsson:2020dhm}.
These studies in particular when they involve also the $\Omega$ baryon will greatly benefit from the large increase in statistics of the 200/pb pp program planned for Run-3/4~\cite{Citron:2018lsq}.

\ifarxiv
\begin{figure}[t!]
\begin{center}
\includegraphics[width=0.85\textwidth]{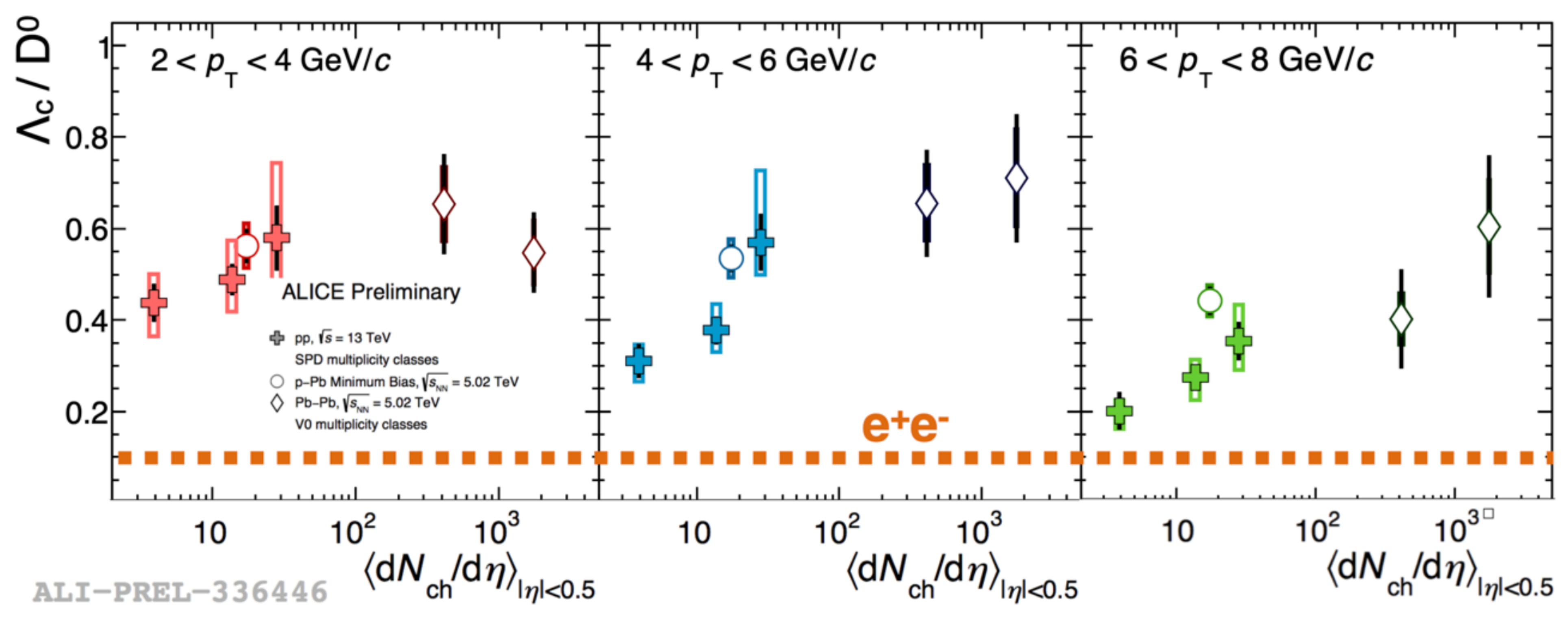}
\caption{\protect\label{fig:lambdacd0}
The $\Lambda^{+}_{c}/D$ ratio versus multiplicity at mid-rapidity measured in several $\pt$-ranges for \pp, \pPb\ and \PbPb\ collisions at $\snn=5.02$ TeV compared to  e$^+$e$^-$ collisions~\cite{Innocenti}.  
}
\end{center}
\end{figure}
\fi

The $\Lambda^{+}_{c}/D$ ratio versus multiplicity at mid-rapidity measured in several $\pt$-ranges for \pp, \pPb\ and \PbPb\ collisions at $\snn=5.02$ TeV~\cite{Innocenti} \ifarxiv shown in \Fig{fig:lambdacd0}\fi exhibits surprising features:
The ratios, in particular for lower $\pT$, smoothly increase with multiplicity from low multiplicity \pp\ to central \PbPb\ collisions.
At the lowest measured multiplicity interval~(below the \pp\ average) the ratio at $2<\pt<4$~GeV/$c$ is already significant~(4 times) larger than that measured in e$^+$e$^-$ collisions, while for the highest multiplicity interval it approaches the value seen in central \PbPb\ collisions.
The result is in qualitative agreement with the ``recombination'' hypothesis saturating already few~($\sim6$) times mean the average \pp\ multiplicity, which requires the presence of a local space-time density.
Alternative mechanisms like implemented in PYTHIA (``color reconnection'') can describe the data (but are not universal).
The new data open up research on hadronization in \pp\ collisions and future \ep~(\eA) colliders.

In \pp\ collisions at 13 TeV, heavier particles up to D-mesons exhibit finite $v_2>0$, while $v_2\approx 0$ for B-mesons~\cite{CMS-PAS-HIN-19-009,Aad:2019aol}.
Unlike at RHIC, where the geometry of the initial state can be engineered through the collisions of proton, deuteron or helium on gold~\cite{PHENIX:2018lia}, at LHC the dependence on geometry can only be more indirectly probed.
Still, measurements of $v_2$ in Z-tagged \pp\ events~(which lead to more ``central'' events), and in photo-nuclear reactions~(which can be regarded as a $\rho$--nucleus collision), reveal finite $v_2>0$~\cite{Aaboud:2019mcw,ATLAS-CONF-2019-022}, while $v_2\approx0$ was found in ee or ep collisions~\cite{Badea:2019vey,ZEUS:2019jya}.
In particular, for those systems where $v_2\approx0$, it would be good to follow up with other measurements like identified particle $\pT$ spectra, mean $\pT$, or yield ratios versus $\pT$ to check for absence of collectivity in observables related to $v_2$.
New calculations for small systems, where initial and final state effects were consistently treated, suggest the dominance of the final state effects beyond ${\rm d}N/{\rm d}\eta>~10$~\cite{Schenke:2019pmk}.
Together with the large set of observables, which in \PbPb\ collisions undoubtedly would be interpreted as QGP signals, this makes it plausible that we indeed observe the sQGP even down to multiplicities just above that of minbias \pp\ collisions at LHC energies.
While in 2015,  there was a large debate whether it is possible~\cite{Loizides:2016tew}, we should now focus on understanding how it is realized in QCD and what it implies.

Although the presence of flow without observable jet quenching does not rule out having a ``mini'' QGP, one of the key questions is the apparent absence of parton energy loss in small systems.
Above a few GeV/$c$, nuclear modification factors measured in minimum bias \pA\ collisions at mid-rapidity are consistent with pQCD calculations including nuclear PDFs~(see references in \cite{Loizides:2016tew}).
Observing a suppression directly in light-particle flavor spectra is complicated due to the selection biased introduced by the ``event-activity categorization''~\cite{Adam:2014qja}.
At high $\pT>10$~GeV/$c$, $v_2$ is understood to reflect the path-length dependence of parton energy loss in \PbPb\ collisions~\cite{Sirunyan:2017pan}.
Therefore, the new data from ATLAS~\cite{Aad:2019ajj}, which demonstrated significant non-zero $v_2$ values up to high $\pT\sim50$~GeV/$c$ in \pPb\ collisions measured using the template fit method, are puzzling.
They can not be consistently interpreted as due to parton energy loss, since the latter would imply $\RpPb$ to be significantly below unity, not seen in data.
Furthermore, the similarity of the $\pt$ shape of $v_2$ between the \pPb\ and \PbPb\ systems provokes the question, whether there is a yet unknown source of $v_2$ in both systems?
In the intermediate region, the anisotropies were found to be larger in minimum-bias than in jet-selected events, and the observed $v_2$  may come from changing the admixture of particles from hard scattering and the underlying event.
At $\pT>10$~GeV/$c$, the ATLAS data essentially report long-range correlation between hard and soft particles on the level of 2\%, which is remarkably similar to that in \PbPb\ collisions, maybe reflecting some source of residual correlation from the likely presence of a di-jet.
It is clear that this question will need to be followed up, maybe by measurements in peripheral \PbPb\ collisions, or by studies using event generators trying to disentangle the various contributions. 

Peripheral \PbPb\ collisions, most of the beam energy systems from STAR, as well as small systems~(\pA\ and \pp collisions) exhibit $v_2$  but no measurable effect from parton energy loss on the $\pT$ spectra~\cite{Acharya:2018njl,Adamczyk:2017nof,Adamczyk:2013gw}.
Hence, one of the next key questions is to study the onset of parton energy loss at RHIC and LHC with lighter ions, like for example in \OO\ or \ArAr\ collisions~\cite{Citron:2018lsq}.
\ifarxiv
Unlike in \pp\ or \pA\ collisions, multiplicity distributions in \AA\ collisions exhibit a pronounced plateau that simplifies the centrality determination, and reduces effects from biases induced by the event selection.
Indeed, it will be very beneficial to perform the same measurements also at $\snn=0.2$~TeV, since the number of MPI per nucleon--nucleon collision will be significantly lower than at LHC energies, while the expected medium properties do not differ greatly.
Furthermore, it may be useful to describe the $\RAA$ data from the RHIC beam energy scan~\cite{Adamczyk:2017nof}, as well as the SPS data~\cite{Alt:2007cd} with models including state-of-the-art initial and final state effects.
\fi

\ifarxiv
\begin{figure}[t!]
\begin{center}
\includegraphics[width=0.48\textwidth]{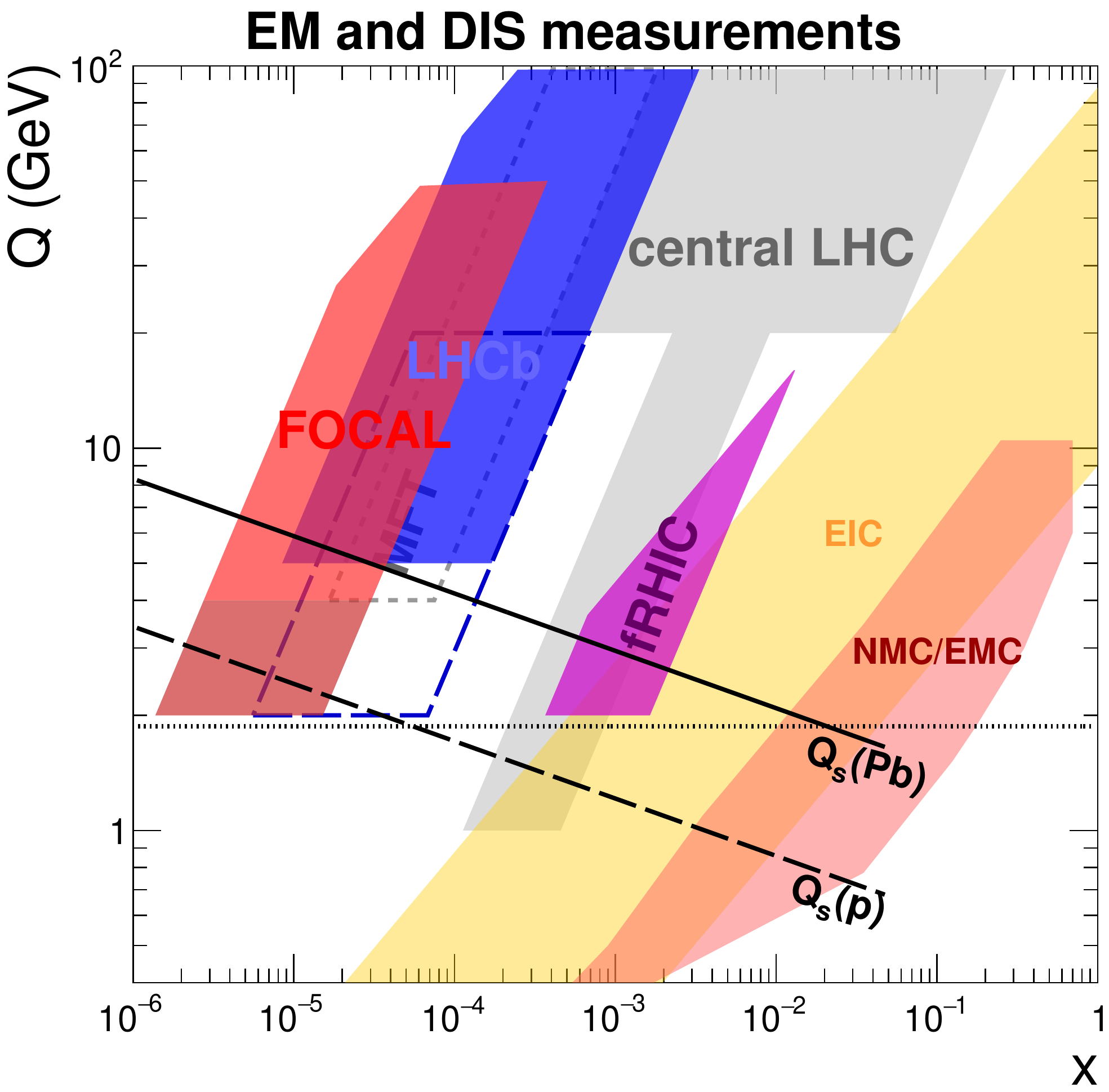}
\caption{\protect\label{fig:xq_coverage} 
Approximate ($x$,$Q$) coverage of various experiments for regions probed by DIS measurements including at the planned EIC, as well as possible future direct photon and Drell-Yan measurements at RHIC and LHC,
The estimated saturation scales for proton and Pb are also indicated.
The horizontal dashed line and the dashed curve indicate the kinematic cuts above which data were included in the nNNPDF fits.
To calculate $x$ and $Q$ the approximate relations $x=2\pt/\snn\exp(-y)$ and $Q=\pT$ was used with $\snn=8.8$~TeV for LHC, and $\snn=0.2$~TeV for RHIC.
Figure from \cite{ALICECollaboration:2719928}.
}
\end{center}
\end{figure}
\fi

\section{The cold-nuclear matter or small-$x$ frontier}
\label{sec:coldmatter}
At small longitudinal momentum fraction $x$ and momentum transfer $Q$, parton dynamics is expected to be affected by non-linear QCD evolution, where the rate of gluon--gluon fusion is in competition with that of  gluon splitting.
In this kinematic regime, the extremely high gluon density may even saturate, possibly leading to the existence of another pre-collision state of matter –-- the so-called colour glass condensate~(CGC)~\cite{Gelis:2012ri}.
The saturation scale, where for a given $x$ the competing processes are in balance, is enhanced in nuclei by a factor  $A^{1/3}$ compared to protons, and hence comparisons between measurements in \pp\ and \pA\ collisions are of particular interest.

\ifarxiv\Figure{fig:xq_coverage} gives an overview of the approximate ($x$,$Q$) coverage of various curent and future experiments for EM or DIS measurements.\fi 
At RHIC and EIC a region down to about $x\sim10^{-4}$ will be probed with the future direct photon and Drell-Yan measurements enabled by the RHIC cold nuclear program~\cite{Aschenauer:2016our}, for which STAR and sPHENIX plan forward upgrades at $2.5<\eta<4$~\cite{phenix,star}.
Shown are also regions probed by nuclear DIS measurements~\cite{Arneodo:1995cs,Arneodo:1995cq,Arneodo:1996ru}, including at the planned EIC~\cite{Accardi:2012qut}.
Due to the large beam energy available at the LHC,  instrumenting the forward region at the LHC enables measurements probing parton densities at small~$x$ of the proton or nucleus, down to $x\sim10^{-6}$, over a large range in $Q^2$.
The LHCb experiment~\cite{Alves:2008zz} is a single-arm spectrometer equipped with tracking and particle-identification detectors as well as calorimeters with a forward angular coverage of about $2<\eta<5$.
The FoCal is high-granularity Si+W electromagnetic calorimeter and metal+scintillator hadron calori\-meter at $3.4<\eta<5.8$ proposed to be installed in ALICE in LS3~\cite{ALICECollaboration:2719928}.
The FoCal and LHCb measurements probe much smaller $x$ than any of the other experiments.
The FoCal will access the smallest $x$ ever measurable until the possible advent of the LHeC~\cite{AbelleiraFernandez:2012cc} or FCC~\cite{Mangano:2018mur}.
Compared to RHIC, the LHC will give access to a significantly larger region of phase space that is potentially affected by parton saturation.
In particular, the region of gluon saturation will extend to $\pT$ values high enough that perturbative QCD should be applicable.

\ifarxiv
\begin{figure}[tbh!]
\begin{center}
\includegraphics[width=0.65\textwidth]{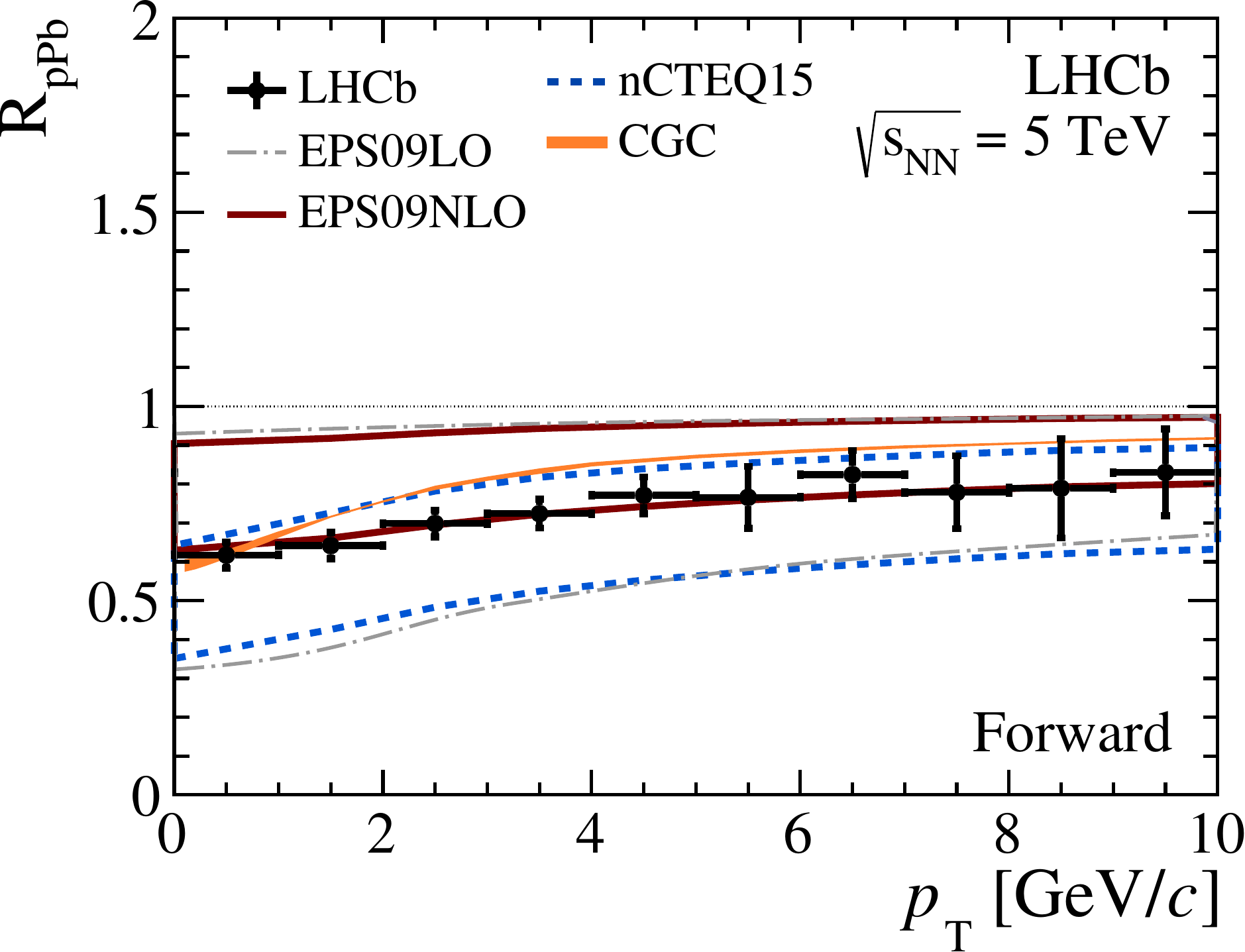}
\caption{\protect\label{fig:dlhcb}
  Nuclear modification factor $\RpPb$ as a function of $\pt$ for prompt $D^0$ integrated over $2.5 < |y^{\ast}| < 4.0$ for $\pt < 6$~GeV/$c$ and $2.5 < |y^{\ast}| < 3.5$ for $6 < \pt < 10$~GeV/$c$ for \pPb\ collisions at $\snn=5.02$~TeV as measured by LHCb compared to theoretical predictions of different pQCD calculations using nuclear PDFs and a recent CGC calculation.
  The figure is adapted from \cite{Aaij:2017gcy}.
}
\end{center}
\end{figure}
\fi

Precise information at forward rapidity probing small $x$ at the LHC is provided by the measurement of prompt D-meson production at $2.5 < y < 4.0$ by LHCb~\cite{Aaij:2017gcy}. 
D-meson production is directly sensitive to the gluon density, since the dominant production process for $c\bar{c}$ production is gluon fusion $gg \rightarrow c\bar{c}$.
The measured nuclear modification factor $\RpPb$ as a function of $\pt$ at forward rapidities\ifarxiv~(\Fig{fig:dlhcb})\fi exhibits that the forward production of prompt D-mesons is suppressed compared to pp collisions, with $\RpPb \sim 0.6$ at low \pt\ and increasing mildly with $\pt$.
The measured suppression is consistent with expectations based on the various calculations using nuclear PDFs or the CGC framework.
The suppression of charm production in the calculations with nuclear PDFs is a direct result of the reduced gluon density at $x \lesssim 10^{-2}$, which is commonly referred to as \textit{gluon shadowing}. 
The calculated values range from $\RpPb$ about $0.3$ to 0.9, reflecting the current uncertainties in the nuclear modification of the small-$x$ gluon density. 
This directly confirms that shadowing at small $x$ is large, and that the data place constraints on nuclear PDFs, as discussed at the conference~\cite{Eskola:2020yfa}.

\ifarxiv
\begin{figure}[thb!]
\begin{center}
\includegraphics[width=0.45\textwidth]{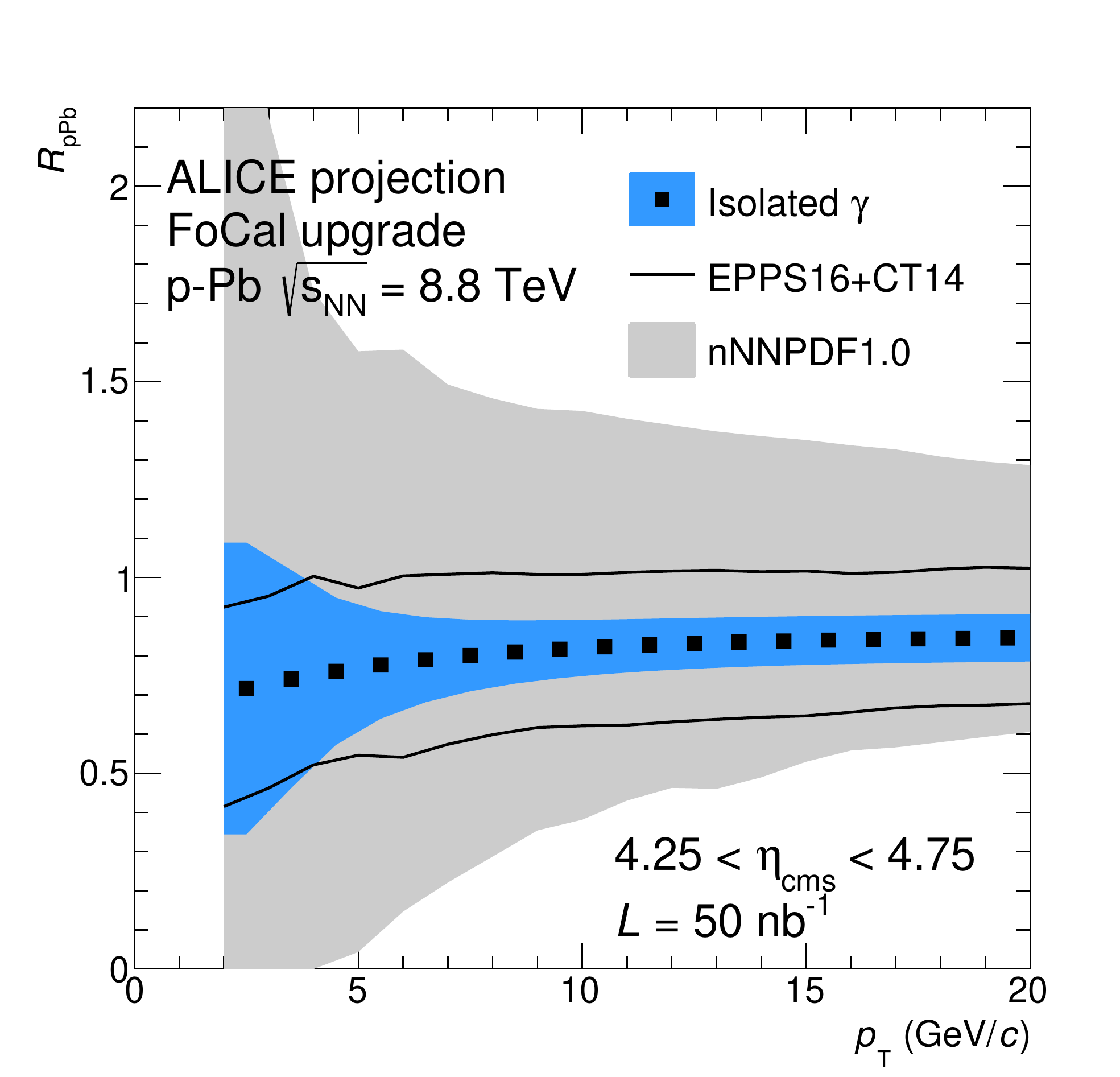}
\includegraphics[width=0.50\textwidth]{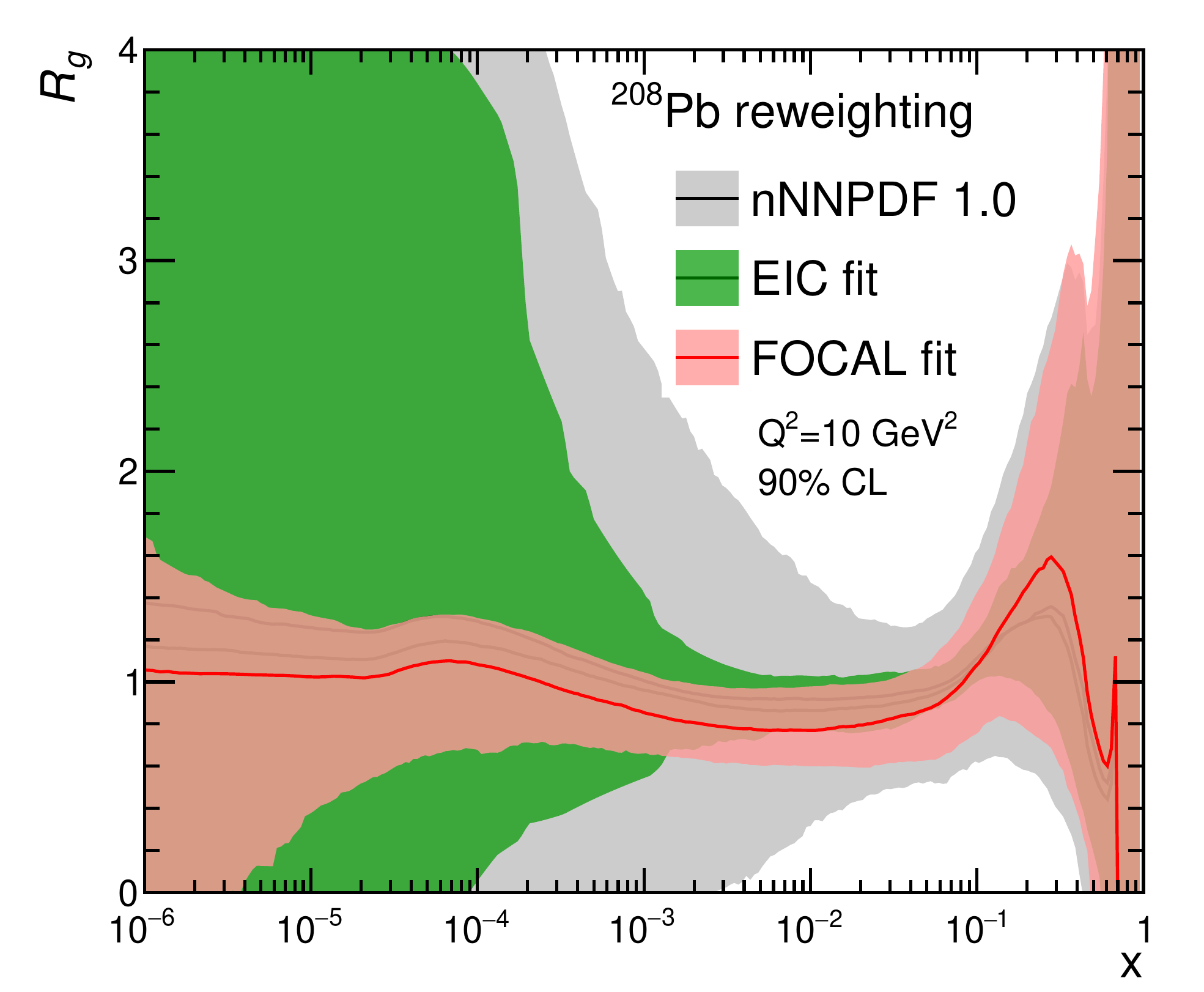}
\caption{\label{fig:gammauncpPb}\label{fig:fitperf}\label{fig:focalperf}
  Left panel: Expected uncertainty for the nuclear modification factor of isolated photons at $\snn=8.8$~TeV measured with FoCal.
  The bands indicate the systematic uncertainty, mostly due to uncertainties on the efficiency and energy scale, as well as the decay photon background determination.
  The current EPPS16 and nNNPDF1.0 uncertainties are indicated by the black line and the shaded band, respectively.
  Right panel: The nuclear modification of the gluon distribution for Pb versus $x$ at $Q^2=10$~GeV$^2/c^2$ for $x>10^{-6}$
  compared between nNNPDF1.0 parameterization and fits to the FoCal pseudo-data (red band) and ``high energy'' EIC pseudo-data (green band).
  90\% confidence-level uncertainty bands are drawn, and the nuclear PDFs are normalized by the proton NNPDF3.1.
  Figures from \cite{ALICECollaboration:2719928}.
}
\end{center}
\end{figure}
\fi

However, a quantitative determination of the amount of gluon shadowing based on hadron production measurements is complicated by the fact that hadronic final state effects~(rescattering) may also play a role in the observations.
Recent forward measurements from LHCb at $\snn=8.16$~TeV include the observations of a stronger suppression of the cross section ratio of $Y(1S)$ to J/$\psi$ from b in \pPb\ compared to \pp\ collisions, as well as stronger nuclear modification of $Y(2S)$ compared to $Y(1S)$~\cite{Aaij:2018scz}.
Both require the presence of additional (final-state) effects to describe the data.
Also, the forward/backward ratio of D-meson production at high $\pT$ in \pPb\ collisions at $\snn=8.16$~TeV~\cite{LHCb:2019dpz} differs from expectation of models and data at 5 TeV.

A clean probe providing direct access to partons are direct photons, since they couple to quarks, and are not affected by final-state nor hadronization effects.
At leading-order more than 70\% are produced by Compton ($qg\rightarrow \gamma q$)  process, hence directly sensitive to the gluon density.
An essential design goal of proposed FoCal~\cite{ALICECollaboration:2719928} is ability to reconstruct $\pi^0\rightarrow\gamma\gamma$ decays at forward rapidity up to large transverse momenta $\pt \sim 20$~GeV/$c$ with high efficiency.
This will enable precise discrimination between direct photons and decay photons, hence enabling to measure direct photons from low transverse momentum up to $\sim 20$~GeV/$c$ at large rapidity.
\ifarxiv
The unique performance of a future isolated photon measurement with FoCal is demonstrated in \Fig{fig:focalperf}.
\fi
An accuracy of about 20\% is expected at $4$GeV$/c$, improving to about 5\% at 10 GeV/$c$ and above, which will strongly constrain nuclear PDFs below $x\sim0.001$, in a region complementary to the EIC.

\section{The ultra-precision near and far future}
\label{sec:future}
\begin{figure}[tbh!]
    \centering
    \includegraphics[width=0.99\textwidth]{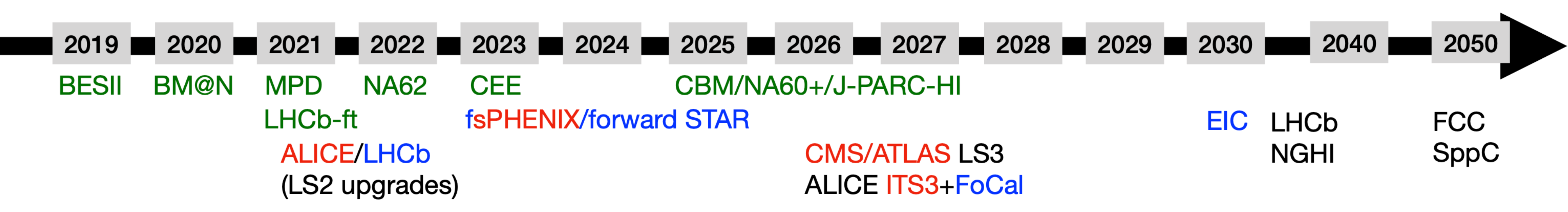}
    \caption{\label{fig:timeline} Timeline of planned and proposed near- and far-future instrumentation for high-energy nuclear physics, grouped into:
      the high-density frontier~(green) with BESII~\cite{Odyniec:2019kfh}, BM@N~\cite{Kapishin:2019wxa}, MPD~\cite{Golovatyuk:2016zps}, LHCb fixed-target~\cite{LHCbCollaboration:2673690}, NA62~\cite{Aduszkiewicz:2309890}, CEE~\cite{Lu:2016htm}, CBM~\cite{Senger:2020pzs}, NA60+~\cite{Dahms:2673280}, J-PARK-HI~\cite{Sako:2015cqa};
      the high-energy frontier~(red) with ALICE LS2 upgrades~\cite{Abelevetal:2014cna,CERN-LHCC-2013-014}, sPHENIX~\cite{Aidala:2012nz}, CMS LS3 upgrades~\cite{Butler:2055167}  and ATLAS LS3 upgrades~\cite{CERN-LHCC-2015-020}, ALICE ITS3~\cite{Musa:2703140}; 
      the small-$x$ frontier~(blue) with LHCb phase-1 upgrade~\cite{LHCb:2018qbx}, forward sPHENIX~\cite{phenix}, forward STAR~\cite{star}, FoCal~\cite{ALICECollaboration:2719928}, EIC~\cite{Accardi:2012qut};
      the ultra-precision future~(black): LHCb phase-2 upgrade~\cite{LHCB2030}, NGHI~\cite{Adamova:2019vkf}, FCC~\cite{Benedikt:2020ejr}, SppS~\cite{CEPC-SPPCStudyGroup:2015csa}.
}
\end{figure}

An overview of the timeline of planned and proposed near- and far-future instrumentation for high-energy nuclear physics, grouped into different categories~(high density, high energy, small-$x$ and ultra-precision future) is given in \Fig{fig:timeline}.
Besides the ongoing experimental program, in particular at RHIC and SPS, new dedicated future experiments and facilities are being built over the next 5--10 years to characterize the phase structure of strongly-interacting matter at high $\mub$~\cite{Galatyuk:2019lcf}.
In addition, at the LHC, data can be taken in fixed-target mode, for example by LHCb with the SMOG2 system~\cite{LHCbCollaboration:2673690}, allowing to probe the freeze-out curve up to $\mub\approx400$~MeV~(see summary in~\cite{Begun:2018efg}).

In 2021 the ALICE LS2 upgrades~\cite{Abelevetal:2014cna,CERN-LHCC-2013-014} to improve the capabilities for rare probes at low $\pt$ will have be completed, which in particular include a new inner tracking system based on MAPS~\cite{Abelev:1625842}, the GEM-based TPC readout~\cite{CERN-LHCC-2013-020}, and the forward muon tracker~\cite{CERN-LHCC-2015-001}.
LHCb, will complete its LS2 upgrades~\cite{LHCb:2018qbx}, with among more minor improvements, a new pixel vertex locator~\cite{Collaboration:1624070} and a new  high-granularity silicon micro-strip planes upstream and scintillating-fibre downstream tracker~\cite{Collaboration:1647400} usable in up to  30--100\% central \PbPb\ collisions.
In 2023, sPHENIX~\cite{Aidala:2012nz} is expected to start operating specifically designed for measurements of hard probes at RHIC.
The major upgrades for CMS~\cite{Butler:2055167}  and ATLAS~\cite{CERN-LHCC-2015-020} are prepared for data taking in 2027 after LS3.
For CMS, they mainly are a new high-granularity pixel and Si-strip tracking system up to $\eta<3.8$~\cite{Collaboration:2272264} with a MIP timing layer for time-of-flight measurements~\cite{CMS:2667167} and a high-granularity calorimeter~\cite{Collaboration:2293646}.
For ATLAS, similarly, they mainly are a new high-granularity pixel and Si-strip tracking system up to $\eta<4$~\cite{Collaboration:2285585}.
In LS3,  the ALICE collaboration will replace the inner-most pixel layers with truly cylindrical layers using thinned, wafer-sized sensors~\cite{Musa:2703140}, which will further improve the secondary vertex finding as well as tracking efficiency and momentum resolution at low $\pt$.

Besides the LHCb after its phase-1 upgrade~\cite{LHCb:2018qbx}, the proposed forward upgrades~\cite{phenix, star} at RHIC and FoCal~\cite{ALICECollaboration:2719928} at the LHC, there is the planned EIC~\cite{Accardi:2012qut} expected to begin data-taking around 2030.
The EIC will be ep and eA polarized~(up to small nuclei) collider dedicated to the study of proton and nuclei structures.
The collider will enable DIS off a proton or a nucleus with a range of $20<\s<140$~GeV and a luminosity of $10^{34}$~cm$^{-2}$s$^{-1}$.
It will allow us to systematically  explore correlations inside protons and nuclei, as well as to study saturation and hadronization with a controllable initial state.
For an overview of the EIC physics program and the connection to heavy-ion physics, given at the conference, see \cite{Hatta:2020nvr}. 

For data-taking at similar timescale as the EIC, the successor of ALICE, the Next Generation Heavy Ion~(NGHI) has been proposed~\cite{Adamova:2019vkf}.
It is designed as a fast, ultra-thin detector with precise tracking and timing, and will provide the ultimate performance for measurements related to  (multi-)heavy-flavor, soft hadrons and thermal radiation over large range in rapidity $\eta\le4$.
Eventually one can add far-forward tracking and particle identification capabilities to access the high net-baryons and/or a far forward calorimeter for ultra-soft-$x$ photons.
In 2030, LHCb will also  have concluded its phase-2 upgrade~\cite{LHCB2030}, after which its tracking detectors will be able to cope with even higher track densities of the HL-LHC.
This will also allow the reconstruction of central \PbPb\ data, and open up the full suite of particle identification provided by the LHCb spectrometer at forward rapidity in\PbPb\ collisions.
Together with ATLAS and CMS, these experiments will provide an incredibly rich, precision high-density QCD and heavy-ion program at the LHC, in parallel to the precision cold nuclear matter program the EIC.
As indicated also in \Fig{fig:timeline}, the far-future perspective is equally bright, given the activities related to the FCC~\cite{Benedikt:2020ejr} including preparations for a heavy-ion program~\cite{Dainese:2016gch} and the CEPC-SppC~\cite{CEPC-SPPCStudyGroup:2015csa}.

\section{Summary}
\label{sec:summary}
Instead of a ``summary of the summary'', let me just point out that ---without any doubt--- we are experiencing the golden age of high-density QCD and heavy-ion physics, with a numerous interesting problems to solve and an extremely bright future in instrumentation ahead.

\section*{Acknowledgements}
I would like to thank the conference organizers for their endless effort to create such a stimulating and successful conference, and for being so extra-ordinary hosts.
Support by the U.S. Department of Energy, Office of Science, Office of Nuclear Physics, under contract number DE-AC05-00OR22725, is greatly appreciated.





\bibliographystyle{elsarticle-num}
\bibliography{clbib}







\end{document}